# AMR-MUL: An Approximate Maximally Redundant Signed Digit Multiplier

Saba Amanollahi[1], Mehdi Kamal[1], Ali Afzali-Kusha[1], Massoud Pedram[2]

[1]School of Electrical and Computer Engineering, University of Tehran
[2]Department of Electrical and Computer Engineering, University of Southern California
s.amanollahi@alumni.ut.ac.ir, {mehdikamal, afzali}@ut.ac.ir, pedram@usc.edu

*Abstract*—**In this paper, we present an energy-efficient, yet high-speed approximate maximally redundant signed digit (MRSD) multiplier (called AMR-MUL) based on a parallel structure. For the reduction stage, we suggest several approximate Full-Adder (FA) reduction cells with average positive and negative errors obtained by simplifying the structure of an exact FA cell. The optimum selection of these cells for each partial product reduction stage provides the lowest possible error, turning this task into a design space exploration problem. We also provide a branch-and-bound design space exploration algorithm to find the optimal assignment of reduction cells based on a predefined constraint (*i.e.*, the width of the approximate part) by the user. The effectiveness of the proposed (Radix-16) multiplier design is assessed under different digit counts and approximate border column. The results show that the energy consumption of the MRSD multiplier is reduced by 7× at the cost of a 1.6% accuracy loss.**

*Keywords*- **Approximate computing; Maximally redundant signed digit multiplier; Approximate FA; Energy reduction; Accuracy.**

## I. INTRODUCTION

To achieve higher throughput in digital systems, the memory bandwidth and computational core performance should be increased. These two parameters are functions of the employed number system to represent data. The conventional binary number system (BNS) is the common system used in nearly all digital devices. Nevertheless, there are some unconventional number systems that are more efficient when doing certain arithmetic operations [1]. Addition and multiplication are two main arithmetic operations used in digital computations. Among the unconventional number systems, the residue number system (RNS) and logarithmic number system (LNS) outperform other number systems for the multiplication operation, while the redundant digit number system (RDNS) is more efficient for the addition operation [2].

Maximally redundant signed digit number system (MRSD) is a version of the RDNS that enables representation of both negative and positive numbers. The digit set of this number system is wider than that for the other redundant number systems leading to a larger dynamic range. The time complexity of the add operation in this number system is $O(1)$ while the multiplication is performed with a time complexity of $O(logn)$ (similar to the BNS) [1]. Due to its superiority for doing the signed addition operation, the redundant signed digit number system has been used as the core of an energy-efficient divider designed in [3]. Indeed, an extensible processor for supporting MRSD-based operations has been presented in [2]. On the negative side, due to need to have redundant bits, more memory is required for storing the data in this representation leading to a higher memory footprint and memory access bandwidth. This makes high radix MRSD an attractive option to be employed in some computational systems. While the delay of a MRSD multiplier is lower than that of a BNS one, more logic cells are needed for the hardware implementation of the MRSD multiplier owing to its larger number of generated partial products. This is also the main reason for the considerably higher energy (power) consumption of the MRSD multiplier compared to the BNS one [2]. In [2], to tackle this problem, the core of MRSD multiplication is implemented by a BNS multiplier where the input MRSD operands were converted to the BNS ones.

On the other hand, many digital systems require higher volume of computations under a given energy budget. Approximate computing has been touted as a promising design paradigm for reducing the power/energy consumption of digital systems while preserving a prespecified level of output accuracy [4]. This approach is applicable to the case of error resilient applications which can tolerate some errors in the output thus making these applications tolerant of imprecise computations. In the approximate computing paradigm, the designer may eliminate some components of the exact design structure (design complexity reduction) to reduce the power consumption and possibly improve the speed [5]. In fact, approximate computing has been widely used in the BNS arithmetic units, including multipliers which are critical and highly used components in digital computing (see, *e.g.*, [5-9]). For parallel multipliers, the common approximation method is the structure complexity reduction, which utilizes approximate compressors/reduction cells [8].

While there are many proposed BNS multipliers in the literature, to the best of our knowledge, there is not any published approximate MRSD multiplier with the aim of reducing energy consumption. In this work, we present an energy efficient yet high speed approximate parallel MRSD multiplier structure (named AMR-MUL). For this structure, we propose to use some approximate FA cells as the reduction cells. The proposed FAs can have either positive or negative average errors. Indeed, the possibility to have approximate FAs with either positive or negative average error emanates from the fact that there are two types of bits with different polarity in the employed MRSD encoding. The ability of selecting a proper combination of these approximate FAs provides us with opportunity of reducing the overall error leading to low multiplication output error. To accomplish this design objective, a design space exploration algorithm based on branch-and-bound algorithm is presented. By defining proper bounds in the proposed algorithm, some of the search tree branches are pruned leading to fast exploration. In the considered bounding This algorithm selects the proper set of proposed approximate FAs during the partial product reduction stage of the multiplication operation. It should be noted that most of the proposed compressors in the prior work tend to produce negative errors (*e.g.*, [8, 10]) leading to a non-Gaussian error distribution, which is not good for producing an accurate signed multiplication result. The AMR-MUL is evaluated under different approximate part widths and its design parameters are compared with those of some state-of-the-art approximate BNS multipliers.

The rest of this paper is organized as follows. In Section II, the considered maximally redundant signed digit number system and the MRSD multiplier structure are briefly discussed. The proposed MRSD multiplier, approximate FAs, and suggested design space exploration algorithm are explained in detail in Section III. In Section IV, the proposed MRSD is assessed and, finally, the paper is concluded in Section V.

## II. Background

### A. Maximally redundant signed digit number system

The input operands and the result of the proposed approximate multiplier are in the maximally redundant signed digit number system. In this work, without loss of generality, we suggest a radix-16 approximate MRSD multiplier. In the radix-16 of this number system, each digit is represented by 5 bits in 2's complement format. We use the encoding representation which is introduced in [11]. Figure 1.a shows the $K^{th}$ digit of this encoding, which digit set is $[−16,15]$. There are two types of bits with the opposite polarity in this encoding i.e., negabit (*i.e.*, negative bits in $\{−1,0\}$) and posibit (*i.e.*, conventional positive bits in $\{0,1\}$). The negabit is placed in the most significant bit (the white circle in the $K^{th}$ digit illustrated in Figure 1.a) whereas the other bits are posibits in the digit (the black circles in the $K^{th}$ digit illustrated in Figure 1.a). In this redundant representation, the negabit of the $K^{th}$ digit has the same weight as the least significant bit of the $(K + 1)^{st}$ digit. The weight and values of the negabit and the two posibits of the $K^{th}$ digit are shown Figure 1.a.

### B. MRSD Multiplier

The general structure of the MRSD multiplier is similar to that of the BNS multiplier. The structure of the proposed approximate multiplier is based on Wallace tree. An exemplary structure for radix-16 MRSD multiplier with 1-digit input operands each is given in Figure 1.b. In this number system, the partial products (PPs) are generated based on the polarity of the input bits using the following bit-level operations [2]:

● × ● = ● AND ● = ●

● × ○ = ● OR NOT(○) = ○

○ × ● = NOT(○) OR ● = ○

○ × ○ = ○ NAND ○ = ●

In the partial product reduction (PPR) stage, the number of PP rows is reduced to two by using reduction cells. Since, in this work, we only use FA and HA as the reduction cells, Figure 1.b shows the PPR stages when FA and HA are used. In the MRSD multiplier, the final merge stage is not required which leads to lower delay compared to that of the BNS multiplier. The output of the PPR stage, however, is not in the format of the input operands. Thus, by considering negabits (with value of 1) and posibits (with value of 0) in the specific positions and using XOR gate in some binary positions, the result of the multiplication is converted to an intermediate binary signed digit (BSD) format. The details of converting to the BSD format is provided in [11]. Then, by using 4-bit adders, the multiplication result, in the format of the radix-16 MRSD, is generated. Note that the width of the result of the $N × N$ MRSD multiplication is $2N + 1$ digits. In this case, by using a few logic gates, the extra digit may be removed, and the value of the removed digit could be represented by updating the most significant negabit.

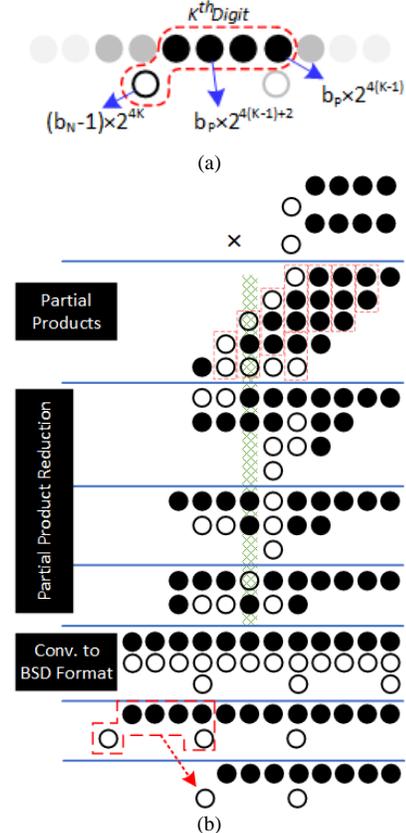

Figure 1. a) The $K^{th}$ radix-16 redundant digit in $[−16,15]$, and b) the internal structure of the of the MRSD multiplier.

## III. PROPOSED APPROXIMATE ARCHITECTURE

The general architecture of the proposed approximate MRSD multiplier is similar to the one depicted in Figure 1.b. In this structure, the exact HA and exact and proposed approximate FAs are employed in the partial product reduction stages. We partition each reduction stage from the $b^{th}$ column of partial products (*e.g.*, the column highlighted by green box in Figure 1.b) into two parts, approximate (from first column to $(b-1)^{st}$ column) and exact (from $(b+1)^{st}$ column to the last column). In the approximate part, for reducing the bit levels, exact HA and proposed approximate FAs are used, while for the exact part the exact HA and FA are employed. For the border column ($b^{th}$ column), exact HA, exact FA, and proposed approximate FAs are utilized. To minimize the accuracy loss, we suggest FAs with positive and negative errors. Thus, choosing them to be used in each column of the partial product reduction stage could be considered as a design space exploration problem. Obviously, the objective for this optimization problem should be the minimization of the output error. For this, we suggest a branch-and-bound algorithm. We considered an approximate border column for all the PPR stages to reduce the error which is imposed by the approximate part. In the following two subsections, the details of the proposed approximate FAs as well as the proposed design space exploration algorithm are described.

### A. Proposed Approximate FAs

Based on the combination of the input bits polarities of the FA employed in the reduction stage, four combinations of the polarities for the output bits are possible which are considered in suggesting our six approximate FAs. The structures, which are illustrated in Figure 2, have similar area usage. In the following, approximate FAs are described in detail:

1) **FA$_{PP}$:** We suggest this approximate structure *when both the sum and carry bits are posibits*. The average error of this structure is $+0.25$. The exact and approximate output values of the exact and proposed approximate FAs are illustrated in Figure 2. Since in this work we considered Radix-16 MRSD representation, the number of posibit partial products is considerably more than that of negabits, making FA$_{PP}$ as the dominant assigned approximate FAs.

2) **FA$^1_{PN}$ and FA$^2_{PN}$:** We propose these two structures *when the output sum bit is negabit and the output carry bit is posibit*. These structures have the average errors of $+0.25$ and $-0.5$, respectively. The motivation for suggesting these structures is as follows. Owing to the larger number of posibits in the columns and the use of the FA$_{PP}$ for the reduction in most cases, the average error of the column reduction becomes positive. On the other hand, due to the smaller number of negabits, the FA$_{PN}$ is employed less frequently compared to the FA$_{PP}$. Thus, we suggest FA$^2_{NP}$ with large negative error to somehow compensate for the positive error of the FA$_{PP}$. Along the same line of reasoning, we suggest FA$^1_{PN}$ structure with small positive error which could be used in the most significant bits where the number of negabits are more compared to that of the least significant bits.

3) **FA$^1_{NP}$ and FA$^2_{NP}$:** We present two structures *when the output sum bit is posibit and the output carry bit is negabit*. These structures have the average errors of $-0.25$ and $+0.5$, respectively. Let us elaborate on the motivation behind the suggestion of these approximate FAs. Due to the small number of the negabits, the chance of assigning FA$_{NP}$ compared to that of FA$_{PP}$ is low. The use of a structure with a low negative error for compensating the error of FA$_{PP}$ and FA$^1_{PN}$ would be desired. Also, the reason for proposing a structure with large positive error is to balance the negative error imposed by the other suggested approximate FAs.

4) **FA$_{NN}$:** We suggest one structure for the case *when both output bits are negabits*. The average error of this approximate FA is $-0.25$. Since the chance of assigning this structure is low (due to the smaller number of negabits), we propose a structure with a small negative error to reduce the

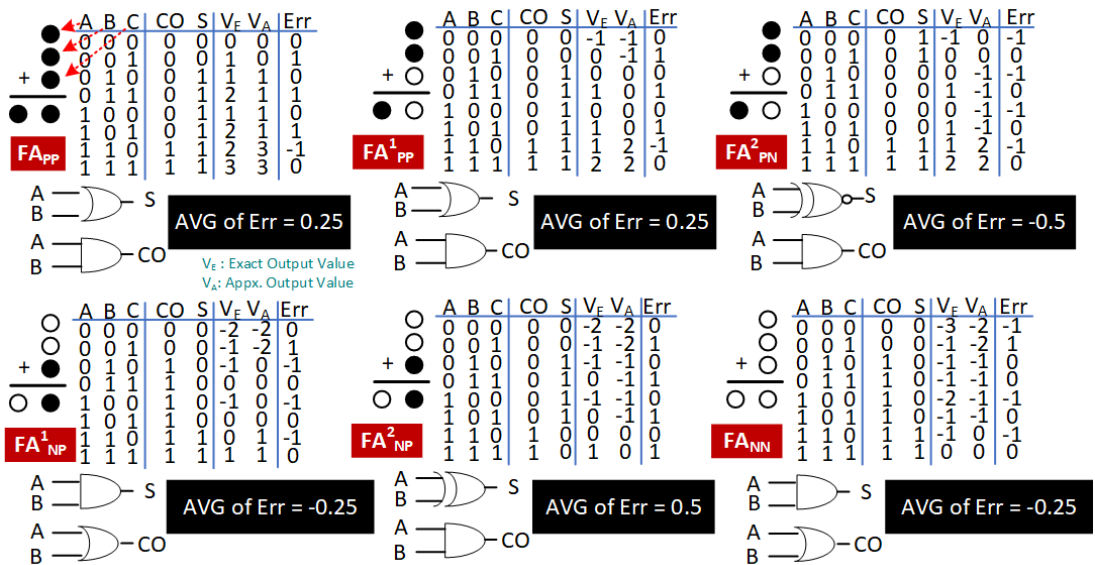

Figure 2. The internal structures of the proposed approximate FAs and their corresponding truth tables and average errors.

```
function DSE_FA_ Assign (pos_cnt, neg_cnt, Err)
#----Checking the bounds-------------
1:   FA_cnt = (pos_cnt + neg_cnt) % 3
2:   if abs(Lowest_Err – Err) < FA_cnt * 0.5
3:      return [], max_number
4:   if (neg_cnt == 0) & ((Err < 0) | (abs(Err + FA_cnt × FA_PP_err ) >
Lowest_Err))
5:      return [], max_number
6:   if (pos_cnt == 0) & ((Err > 0) | (abs(Err + FA_cnt × FA_NN_err ) >
Lowest_Err))
7:      return [], max_number
#----Exploring the branches of the search space tree ----------------------
8:   if (FA_cnt == 0)
9:      if (Err < Lowest_Err)
10:        Lowest_Err = Err
11:     return [], Err
12:  E =max_number
13:  if (pos_cnt >= 3) #Assigning a FA_PP
14:     A[0], E[0] = DSE_FA_Assign (pos_cnt – 3, neg_cnt, Err + FA_PP_err)
15:  if (pos_cnt >= 2) & (neg_cnt >= 1) #Assigning a FA$^1_{PN}$
16:     A[1], E[1] = DSE_FA_ Assign (pos_cnt – 2, neg_cnt – 1, Err + FA$^1_{PN\_err}$)
17:  if (pos_cnt >= 2) & (neg_cnt >= 1) #Assigning a FA$^2_{PN}$
18:     A[2], E[2] = DSE_FA_ Assign (pos_cnt – 2, neg_cnt – 1, Err + FA$^2_{PN\_err}$)
19:  if (pos_cnt >= 1) & (neg_cnt >= 2) #Assigning a FA$^1_{PN}$
20:     A[3], E[3] = DSE_FA_ Assign (pos_cnt – 1, neg_cnt – 2, Err + FA$^1_{NP\_err}$)
21:  if (pos_cnt >= 1) & (neg_cnt >= 2) #Assigning a FA$^1_{PN}$
22:     A[4], E[4]= DSE_FA_ Assign (pos_cnt – 1, neg_cnt – 2, Err + FA$^2_{NP\_err}$)
23:  if (neg_cnt >= 3) #Assigning a FA_NN
24:     A[5], E[5] = DSE_FA_ Assign (pos_cnt , neg_cnt – 3, Err + FA_NN_err)
25:  index = min(E)
26:  return [index, A[index]], E[index]
```

Figure 3. The pseudo-code of the proposed branch-and-bound design space exploration algorithm.

positive error induced by the approximate FA with positive average errors.

### B. The Propsoed Design Space Exploration Algorithm

As mentioned before, due to the availability of a set of approximate FAs with positive and negative average errors and also, the existence of two types of bits in each column, assigning proper FAs in each column could be considered as an optimization problem. Thus, we suggest a branch-and-bound assignment algorithm whose pseudo-code is given in Figure 3. This algorithm should be called for each column of the approximate part to reduce the number of rows in each partial product reduction stage. Its inputs are the number of the posibits ($pos\_cnt$) and negabits ($neg\_cnt$) in the corresponding column and the total error of the approximate FA assignment process in the corresponding column of the previous PPR stage and the corresponding estimated average error of the assignment process in the previous column of the current PPR stage. The proposed procedure is called recursively to traverse all branches of the search space tree. Each branch shows the assignment of one of the proposed approximate FAs. The assignment of an approximate FA depends on the available posibits and negabits in the column where, after assigning a FA in a search space branch, the proposed algorithm updates the number of negabits and posibits. By moving to a branch, the amount of the error is updated based on the assigned approximate FA.

While completely traversing branches of the search tree leads to the optimal solution, this is not feasible for large multipliers where there is a large number of rows in each PPR stage. In addition, finding the best solution does not necessarily require searching the full branches in the algorithm. Thus, in the proposed algorithm, we bound the search space through obtaining the minimum possible error when traversing a branch. More specifically, when the estimated average error of a branch becomes larger than that of the best determined assignment, the algorithm does not continue the search in the corresponding branch. We consider three bounding cases that are given below:

1) The absolute difference between the current obtained error and the best assignment error may not be compensated by assigning the remaining approximate FA. This bounding case is independent from the available bit polarities and occurs when the error is considerably large.

2) Only the posibits remain and the current error may not be compensated by assigning FA_PP. The possibility of encountering this bounding case is much more than the other cases.

3) Only the negabits remain and the current error may not be compensated by assigning FA_NN. Since the number of the negabits in partial products are limited, the chance of encountering this bounding case is very low when compared to the other cases.

These bounding cases do not prevent the algorithm from reaching to the best assignment. As mentioned before, the function *DSE_FA_Assign* is employed for the row reduction of the approximate part columns. For the border column (*i.e.*, $b^{th}$ column), a function similar to *DSE_FA_Assign* is employed while in its branching phase, exploring the exact FA assignment is considered in addition to the proposed approximate ones.

## IV. RESULTS AND DISCUSSION

In this section, the efficacy of the proposed Radix-16 AMR-MUL versus different input operand bit widths is assessed using 45nm NanGate technology [12]. We considered 2-, 4- and 8-digit input operands for AMR-MUL. While having larger dynamic range, they may be considered almost equivalent to the 8-, 16- and 32-bit binary number system in terms of bitwidth, respectively. For extracting design parameters, Synopsys Design Compiler is employed and the considered designs are synthesized for the maximum frequency constraints. Also, to assess the multiplier accuracy when approximate FAs are used, the 2-, 4- and 8- digits designs are simulated by injecting 50K, 500K and 1M random inputs. In the following subsections, first, the accuracy and design parameters of the AMR-MUL under different approximate border columns are studied. Next, the accuracy and design parameters of the AMR-MUL are compared with those of some state-of-the-art approximate BNS multipliers.

### A. Accuracy and Design Parameters of the AMR-MUL

The accuracy results of the proposed structures are reported in Table I. It contains the mean relative error distance (MRED), mean absolute relative error distance (MARED) and normalized mean error distance (NMED) of the AMR-MUL with different border column positions. As was expected, increasing the approximate part width decreases the accuracy quantified by all the considered error metrics. Due to the normal distribution ($\mu \approx 0$) of the relative error of the AMR-MUL, its MARED is larger than the MRED. Also, enlarging the width of the approximate part increases the difference between the MARED and MRED. This originates from the fact that the RE distribution is widened by increasing the approximate part width.

TABLE I. MRED, MARED AND NED OF THE AMR-MUL UNDER DIFFERENT BORDER BIT POSITION.

| | Appx. Bor. Col. | 6 | 7 | 8 | 9 | 10 |
|---|---|---|---|---|---|---|
| 2 Digits | MRED | 1.29E-02 | -2.12E-03 | 2.03E-03 | 5.70E-04 | -4.57E-02 |
| | MARED | 2.98E-02 | 4.37E-02 | 1.06E-01 | 2.68E-01 | 5.97E-01 |
| | NMED | 4.00E-04 | 5.98E-04 | 1.25E-03 | 3.34E-03 | 7.34E-03 |
| | Appx. Bor. Col. | 12 | 15 | 18 | 21 | 24 |
| 4 Digits | MRED | 1.31E-04 | 2.35E-03 | 1.18E-02 | 6.90E-02 | 1.76E-01 |
| | MARED | 2.71E-04 | 3.88E-03 | 2.50E-02 | 1.51E-01 | 5.33E-01 |
| | NMED | -1.00E-06 | -7.00E-06 | -7.70E-05 | -2.76E-04 | -3.43E-03 |
| | Appx. Bor. Col. | 45 | 48 | 50 | 53 | 55 |
| 8 Digits | MRED | 1.06E-04 | 5.52E-04 | 2.71E-03 | 3.90E-02 | -1.97E-02 |
| | MARED | 9.29E-04 | 7.09E-03 | 1.61E-02 | 1.58E-01 | 5.18E-01 |
| | NMED | 3.00E-06 | 1.50E-05 | 5.60E-05 | 4.34E-04 | 2.36E-03 |

TABLE II. DESIGN PARAMETERS OF THE AMR-MUL UNDER DIFFERENT BORDER BIT POSITION.

| | Appx Bor. Col. | Exact | 6 | 7 | 8 | 9 | 10 |
|---|---|---|---|---|---|---|---|
| 2 Digits | Delay (ns) | 0.73 | 0.72 | 0.71 | 0.71 | 0.71 | 0.69 |
| | Power (mW) | 0.87 | 0.84 | 0.75 | 0.59 | 0.50 | 0.37 |
| | Energy (pJ) | 0.63 | 0.61 | 0.54 | 0.42 | 0.36 | 0.25 |
| | Area (µm²) | 1263 | 1297 | 1145 | 972 | 844 | 764 |
| | Appx. Bor. Col. | Exact | 12 | 15 | 18 | 21 | 24 |
| 4 Digits | Delay (ns) | 1.04 | 1.03 | 1.00 | 0.94 | 0.91 | 0.73 |
| | Power (mW) | 4.67 | 3.41 | 2.85 | 2.32 | 1.49 | 1.03 |
| | Energy (pJ) | 4.85 | 3.51 | 2.85 | 2.18 | 1.36 | 0.75 |
| | Area (µm²) | 5408 | 4120 | 3617 | 3243 | 2358 | 2167 |
| | Appx. Bor. Col. | Exact | 45 | 48 | 50 | 53 | 55 |
| 8 Digits | Delay (ns) | 1.23 | 1.11 | 1.05 | 1.00 | 0.95 | 0.95 |
| | Power (mW) | 16.91 | 4.07 | 3.23 | 2.93 | 2.07 | 1.52 |
| | Energy (pJ) | 20.80 | 4.51 | 3.39 | 2.93 | 1.96 | 1.44 |
| | Area (µm²) | 18330 | 6815 | 6207 | 5794 | 5085 | 4583 |

Additionally, as the figures indicate, by increasing the input operands digits, the accuracy is improved. This may be attributed to the fact that for larger multiplier widths, there are more rows (specially, in the beginning stages) in the columns of the PPR stage. This provides us with more opportunity of finding the set of the proposed approximate FAs with lower errors during the design space exploration using the proposed algorithm. In short, for a given error constraint, for larger AMR-MUL structures, we can consider wider approximate part.

Table II shows the delay, power, energy and area usage of the proposed AMR-MUL under different border bit positions compared to the exact MRSD multiplier. As the results show, by widening the approximate part, all the parameters are improved, especially, the power (energy) and area that are reduced considerably. In the case of the 2-digit (8-digit) AMR-MUL, by considering the approximate border position of 8 (50), the delay, power, energy and area are improved by 2%, 32%, 34% and 23% (18%, 82%, 85% and 68%), respectively. Also, by widening the approximate part from border column of 6 to 10 (45 to 55) in the case of 2-(8-)digit AMR-MUL, the delay, power, energy and area are reduced by 4%, 56%, 58% and 41% (14%, 62%, 68% and 32%), respectively.

Finally, Figure 5 illustrates the percentages of the FA types usage in the two AMR-MUL structures. As we have expected, $FA_{PP}$ has been employed more than the other FAs. On the other hand, $FA^2_{NP}$ has been utilized lower than the others which is due to its average positive error. However, as the provided plots show, by increasing the width of the multiplication and the approximate part, its usage is increased. The amount of exact FA utilization depends on the exact part width of the AMR-MUL. Since in both considered structures, the exact part is small, the exact FA usage is low.

### B.  Comparing AMR-MUL with Approximate Binary Multipliers

In this subsection, the design parameters and accuracy of the proposed AMR-MUL are compared with those of some state-of-the-art BNS approximate multipliers. The approximate multipliers proposed in [9] (including mul8s_KVA, mul8s_L1G, mul16s_GAT, and mul16s_HFZ), approximate multipliers proposed in [8] (including C-FULL and C-N configurations where each of them has a standard and hybrid structures), LETAM [13] (denoted by LETAM($X,Y$) where $X$ and $Y$ are the bit lengths of the truncated input operands of its internal main addition and multiplication, respectively), TOSAM [14] (denoted by TOSAM($X,Y$) where $X$ and $Y+1$ are the bit lengths of the truncated input operands of its internal main addition and multiplication, respectively), DRUM [15] (denoted by DRUM($X$) where $X$ is the bit length of the rounded input operands), AS-RoBA [16] and DQ4:2C4 [10].

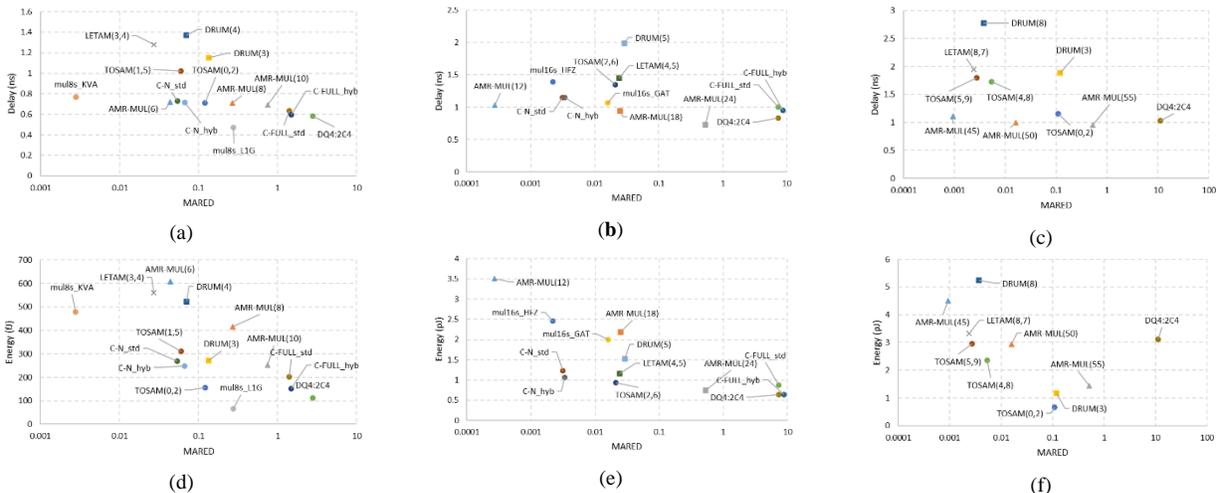

Figure 4. Delays of the a) 2-digit AMR-MUL and some 8-bit approximate multipliers, b) 4-digit AMR-MUL and some 16-bit approximate multipliers, and c) 8-digit AMR-MUL and some 32-bit approximate multipliers respect to their MAREDs. Energies of the d) 2-digit AMR-MUL and some 8-bit approximate multipliers, e) 4-digit AMR-MUL and some 16-bit approximate multipliers, and f) 8-digit AMR-MUL and some 32-bit approximate multipliers respect to their MAREDs.

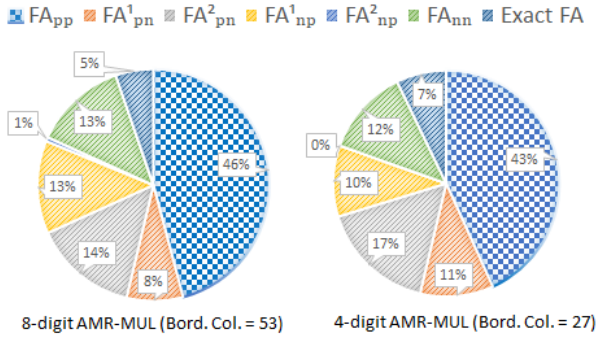

Figure 5. The percentages of the employed FA types in two AMR-MULs.

The design parameters of 2-, 4- and 8-digit AMR-MULs with those of the 8-, 16- and 32-bit BNS approximate multipliers, respectively, are compared in Figure 4. The dynamic range of the $X$-digit AMR-MUL is more than its corresponding considered BNS approximate multipliers. For example, the dynamic range of the 2-digit MRSD (8-bit BNS) is [−272,255] ([−128,127]). Also, since the 32-bit approximate multipliers proposed in [9] and [8] are not available, they are not included in the comparative study.

As the results show, for similar accuracies, the AMR-MUL in all the cases, except for the case of large MARSD errors, has a lower delay. As mentioned before, the higher speed is the main advantageous feature of the MRSD arithmetic units. While some BNS approximate multiplier structures (such as LETAM, TOSAM, and DRUM) have larger delays than that of the exact BNS multiplier, the delay of the AMR-MUL is always smaller or equal to the exact MRSD multiplier. Note that the delays of the exact 8-, 16- and 32-bit exact BNS multipliers are 0.89ns, 1.22ns and 1.65ns, respectively. By increasing the width of the input operands, the superiority of the AMR-MUL in delay compared to approximate BNS multipliers is improved thanks to larger numbers of partial products in the PPR stage, and hence, more opportunity for using different proposed approximate FA to lower the error.

On the other hand, as was expected, due to the larger number of partial products, in most cases, the energy consumption of the AMR-MUL is larger than that of the BNS approximate multipliers. However, when the width of the operands and maximum MARED constraint increase, the energy consumption of the proposed multiplier comes into the range of those of BNS approximate multipliers. Also, it should be noted that while the energy consumption of the exact MRSD multiplier is higher than that of the exact BNS multiplier, the energy consumption of the AMR-MUL is considerably lower than the exact BNS multiplier. The energy consumption of the exact 8-, 16- and 32-bit BNS multipliers are 0.24pJ, 2.6pJ and 17.5pJ, respectively. Finally, though the delay of the proposed AMR-MUL is not lower than those of some of the BNS approximate multipliers at large MARED values, the relative error distribution of AMR-MUL had a Gaussian distribution. In the case of the mul8_L1G which is a BNS multiplier, its relative error may not follow a Gaussian distribution very well (see Figure 6).

## V. Conclusion

In this paper, we proposed an approximate MRSD multiplier structure. The proposed structure was based on parallel multipliers, and thus, we suggested a set of approximate FAs (as the reduction cells) with either positive or negative average error. In addition, to proper assigning the proposed approximate FAs in the PPR stages to reach lower output error, we suggested a branch-and-bound design space exploration algorithm. The results show that the proposed structure reduced the energy consumption of the MRSD multiplier considerably.

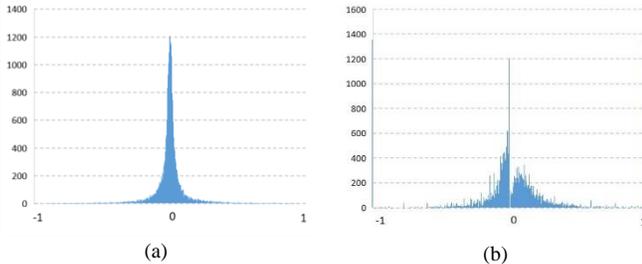

(a)         (b)

Figure 6. The relative error distributions of the 2-digits (a) AMR-MUL and (b) mul8s_L1G in the range of 1 and -1.